\documentclass[10pt]{article}
\usepackage[square,authoryear]{natbib}
\usepackage{graphicx}
\usepackage{stackrel}
\usepackage{cancel}

\usepackage{epstopdf,framed}
\usepackage{pdfsync}
\usepackage{amscd}
\usepackage{mathrsfs}

\RequirePackage{color}
\usepackage{amsmath,bm}
\usepackage{amssymb}
\usepackage{amsfonts}
\usepackage{url}

\begin{document}
\newtheorem{remark}[theorem]{Remark}

\title{Non-isothermal diffusion in interconnected discrete-distributed systems: a variational approach\vspace{1cm}}

\date{}
\author{\hspace{-1cm}
\begin{tabular}{cc}
Fran\c{c}ois Gay-Balmaz & Hiroaki Yoshimura
\\ CNRS, LMD, IPSL & School of Science and Engineering
\\  Ecole Normale Sup\'erieure & Waseda University 
\\   24 Rue Lhomond 75005 Paris, France &  Okubo, Shinjuku, Tokyo 169-8555, Japan
\\ gaybalma@lmd.ens.fr & yoshimura@waseda.jp \\
\end{tabular}\\\\
}
\maketitle
\vspace{-0.3in}

\newcommand{\todo}[1]{\vspace{5 mm}\par \noindent
\framebox{\begin{minipage}[c]{0.45 \textwidth}
\tt #1 \end{minipage}}\vspace{5 mm}\par}

\begin{abstract}
Motivated by compartmental analysis in engineering and biophysical systems, we present a variational framework for the nonequilibrium thermodynamics of systems involving both distributed and discrete (finite dimensional) subsystems by specifically using the ideas of interconnected systems. We focus on the process of non-isothermal diffusion and show how the resulting form of the entropy equation naturally yields phenomenological expressions for the diffusion and entropy fluxes between two compartments, which results in generalized forms of Robin type interface conditions.

\end{abstract}

\tableofcontents


\section{Introduction}

Compartment modeling is a standard technique in the study of the dynamics of engineering or biophysical systems. In the simplest situations, each compartment can be efficiently described by homogeneous thermodynamic quantities, thus giving rise to a finite dimensional interconnected thermodynamic system. Some studies, for instance in intracellular dynamics, however require certain compartments to be considered as spatially distributed subsystems, while others can still be treated via a finite dimensional description. In this case, one needs to describe the thermodynamics of an interconnected discrete-distributed system, where the idea of {\it interconnections}, originally developed by \cite{Kr1963} plays a key role in modeling such complicated systems.

In this paper we present a unified variational framework for the nonequilibrium thermodynamics of such interconnected systems based on the variational formulation for nonequilibrium thermodynamics developed in \cite{GBYo2016a,GBYo2016b,GBYo2019}. This formulation extends the Hamilton principle of classical mechanics to incorporate irreversible processes in both discrete (finite dimensional) and continuum systems. It consists of a critical action condition subject to two types of constraints:
\begin{itemize}
\item[(i)] a phenomenological constraint imposed on the critical curve;
\item[(ii)] a variational constraint imposed on the variations to be considered in the action functional.
\end{itemize}
The phenomenological constraint is directly constructed from the expression of the entropy production written in terms of the thermodynamic fluxes and forces involved in the systems (\cite{DGMa1969}). The phenomenological and variational constraints are systematically related thanks to the introduction of the concept of thermodynamic displacement associated to an irreversible process, defined such that its time rate of change equals the thermodynamic force of the process, \cite{GBYo2016a,GBYo2016b,GBYo2019}. From a mathematical point of view, this variational formulation appears as a nonlinear generalization of the Lagrange-d'Alembert principle of nonholonomic mechanics.

An important property of this variational formulation for the present work is the similar form that it takes for both discrete (finite dimensional) and continuum systems, which allows to naturally treat interconnected discrete-distributed systems. 
%
%
%
Our paper is structured as follows. In \S\ref{section1} we review the case of non-isothermal diffusion in interconnected discrete systems. In this case, one entropy variable and one molar variable can be associated to each subsystem. The second law naturally yields phenomenological relations given by the discrete analogue to the Fourier and Fick laws, and their cross-effects. In \S\ref{section2} we consider the case of a system involving both discrete and distributed compartments. The same type of variational formulation also applies in this case, both in its finite dimensional and continuum versions. For these systems, the Fourier and Fick laws and their cross-effects naturally appear in their usual continuum version as well as in a discrete form when associated to the transfer between the two compartments, thereby yielding Robin type interface condition. Finally, the case of interconnected distributed compartments is considered in \S\ref{section3}.




\color{black} 

\section{Interconnected discrete systems}\label{section1}

In this section we review from \cite{GBYo2019} the variational formulation for the dynamics of non-isothermal diffusion between discrete interconnected compartments. In such a discrete description it is assumed that each compartment is well-stirred so that spatially uniform thermodynamic quantities can be attributed to each of them.
The systems considered here are useful for the nonequilibrium thermodynamic description of membrane transport in biophysical systems, \cite{OsPeKa1973,KaCu1975}.

\begin{figure}[h]
\begin{center}
\includegraphics[scale=0.45]{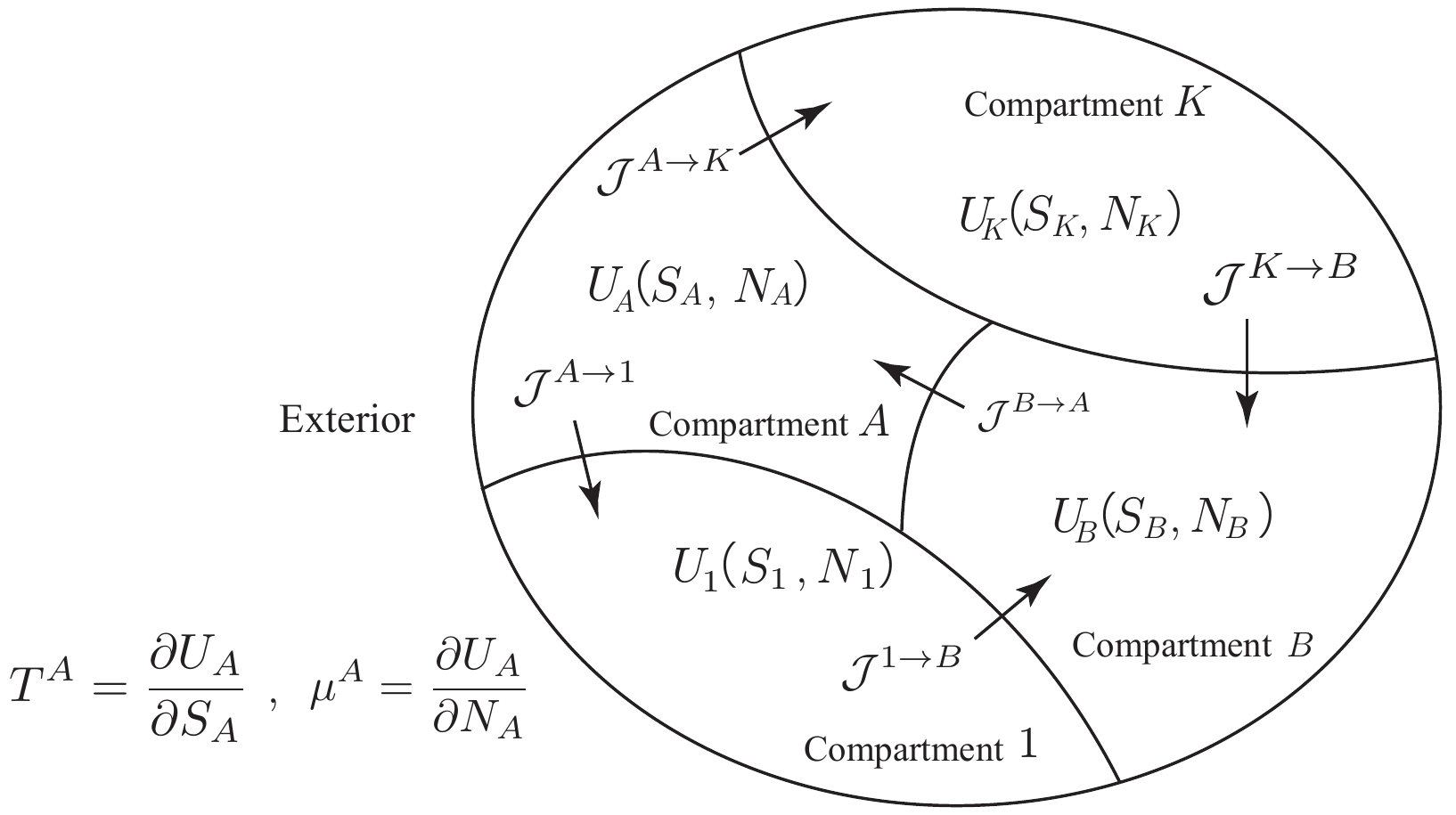}
\caption{\small The heat and matter exchanger}
\label{heated_piston}
\end{center}
\end{figure}

We assume that there is a single species and denote by $N_A$ and $S_A$ the number of moles and the entropy of the species in the compartment $A$, $A=1,....,K$. The internal energies are given as $U_A(S_A, N_A)$, where we assume that the volume of each compartment is constant, see Fig. \ref{heated_piston}. The variational formulation is based on the concept of \textit{thermodynamic displacement} associated to an irreversible process, defined such that its time rate of change equals the {\it thermodynamic force} of the process. In our case, the thermodynamic force are the temperatures $T^A= \frac{\partial U_A}{\partial S_A}$ and  chemical potentials $ \mu ^A= \frac{\partial U_A}{\partial N_A}$, so the thermodynamic displacements are variables $\Gamma ^A$ and $ W^A$ with
\[
\dot \Gamma ^A= T^A \qquad\text{and}\qquad  \dot W^A= \mu ^A.
\]
We also introduce the entropy variables $ \Sigma _A$ whose physical interpretation will be clarified later.

The interfaces are assumed to be diathermal and permeable. We denote by $ \mathcal{J} ^{A \rightarrow B}$ the molar flow rate  from compartment $A$ to compartment $B$ due to diffusion of the species, where $\mathcal{J} ^{A \rightarrow B}=-\mathcal{J} ^{B \rightarrow A}$. We also introduce the fluxes $J_{AB}$ with $J_{AB}=J_{BA}$ and such that $\sum_B J_{AB}=0$, for all $A$, associated to the total power exchange between compartments $A$ and $B$.

\subsection{Variational formulation}
Given a Lagrangian $L$ depending on all the variables, the variational formulation for this class of systems is
\begin{equation}\label{HP_isolated_example2}
\begin{aligned} 
&\delta \int_{t_0}^{t_1} \Big[ L(S_1,...,S_K,N_1,...,N_K)  \\
&\hspace{1.5cm}+ \sum_A\dot W^AN_A + \sum_A\dot \Gamma ^A (S_A- \Sigma _A ) \Big] {\rm d}t=0
\end{aligned} 
\end{equation}
subject to
\begin{equation}\label{constraint_thermo_example2}
\begin{aligned} 
\frac{\partial L}{\partial S_A} \dot  \Sigma _A  &= \sum_BJ_{AB} \dot \Gamma^B + \sum_B\mathcal{J} ^{B \rightarrow A} \dot W^A,\\[2mm]
\frac{\partial L}{\partial S_A} \delta   \Sigma _A  &=\sum_B J_{AB} \delta  \Gamma^B + \sum_B\mathcal{J} ^{B \rightarrow A} \delta  W^A ,
\end{aligned} 
\end{equation}
for $A=1,...,K$, with $ \delta W^A|_{t=t_0,t_1}=\delta \Gamma ^A|_{t=t_0,t_1}=0$.

\subsection{Equations of evolution}

From \eqref{HP_isolated_example2}--\eqref{constraint_thermo_example2}, one directly obtains the conditions
\begin{align*} 
\delta N_A: & \;\;\frac{\partial L }{\partial N_1} =- \dot   W^A &  \;\;\delta S_A:&\;\; \frac{\partial L }{\partial S_A} = - \dot \Gamma ^A \\[2mm]
\delta W^A: & \;\;\dot  N_A = \sum_B \mathcal{J} ^{B \rightarrow A}&  \;\;\delta \Gamma ^A:&\;\; \dot S_A= \dot \Sigma _A+ \sum_B J_{AB},
\end{align*}
for $A=1,...,K$.
Therefore, from $\sum_B J_{AB}=0$ one gets
\begin{equation}\label{equations_discrete}
\left\{
\begin{array}{l}
\vspace{0.2cm}\displaystyle\dot  N_A= \sum_B \mathcal{J} ^{B \rightarrow A}\\[2mm]
\vspace{0.2cm}\displaystyle \frac{\partial L}{\partial S_A} \dot S_A = - \sum_B J_{AB} \left( \frac{\partial L}{\partial S_B}-\frac{\partial L}{\partial S_A}\right)  - \sum_B \mathcal{J} ^{B \rightarrow A} \frac{\partial L}{\partial N_A}. 
\end{array}
\right. 
\end{equation} 
For the particular case of the Lagrangian
\[
L(S_1,...,S_K,N_1,...,N_K)=-\sum_A U_A(S_A,N_A)
\]
given by the sum of the internal energies for each compartments the evolution equations \eqref{equations_discrete} become
\begin{equation}
\left\{
\begin{array}{l}
\vspace{0.2cm}\displaystyle
\dot N_A = \sum_B \mathcal{J} ^{B \rightarrow A}\\[2mm]
T^A\dot S_A = -\sum_B J_{AB}(T^B-T^A) - \sum_ B \mathcal{J} ^{B \rightarrow A} \mu ^A,
\end{array}
\right. 
\end{equation} 
together with $\dot \Gamma ^A= T^A$, $\dot W^A= \mu ^A$, and $ \dot \Sigma _A= \dot S_A$.

\subsection{Balance equations}

From these evolution equations, it consistently follows that the total energy and number of moles are preserved while the total entropy $S= \sum_A S_A$ satisfies
\begin{align*} 
\dot S &= \sum_{A<B} \left( \frac{1}{T^B}- \frac{1}{T^A}\right) J_{AB} (T^B-T^A) +\sum_{A<B}  \left( \frac{ \mu ^B}{T^B}- \frac{ \mu ^A}{T^A}  \right) \mathcal{J} ^{B \rightarrow A},
\end{align*} 
which dictates the choice of phenomenological expressions for $J_{AB}$ and $ \mathcal{J} ^{B \rightarrow A}$ in accordance with the second law of thermodynamics, see \cite{GBYo2019}. For instance, in the linear regime, these expressions read
\begin{equation}\label{pheno_open_system}
\begin{bmatrix}
\vspace{0.2cm}J_{AB}\frac{T^A-T^B}{T^A T^B}\\
\vspace{0.1cm}\mathcal{J}^{B \rightarrow A}\\
\end{bmatrix}= \mathcal{L}_{AB}
\begin{bmatrix}
\vspace{0.2cm}T^B-T^A\\
\vspace{0.1cm}\frac{\mu^{B}}{T^B}-\frac{\mu^{A}}{T^A}\\
\end{bmatrix},
\end{equation}
where the symmetric part of $2\times 2$ matrix $\mathcal{L}^{AB}= \mathcal{L} ^{BA}$ is positive, for all $A,B=1,...,K$.
The entries of these matrices are phenomenological coefficients which may in general depend on the state variables. From Onsager's relation, the matrices $\mathcal{L}^{AB}$ are symmetric for all $A,B$. 
The phenomenological expressions in \eqref{pheno_open_system} describe the discrete version of the Fourier and Fick laws, as well as their cross-effects given by discrete versions of the Soret and Dufour effects.
Assuming that the Lagrangian can be written as the sum $\sum_A L_A(S_A, N_A)$ of Lagrangians associated to each compartment, the balance of energy of the subsystem given by the compartment $A$ is found as
\[
\frac{d}{dt} E_A =\sum_B J_{AB}(T^A- T^B), \quad A=1,...,K.
\]
This allows to relate $J_{AB}$ to the power exchange between compartments $A$ and $B$ as $J_{AB}(T^A- T^B)= P^{B \rightarrow A}$.

The extension of the variational formulation \eqref{HP_isolated_example2}-\eqref{constraint_thermo_example2} to the case of an open system exchanging matter and heat with the exterior is presented in \cite{GBYo2018a}, while the case of reacting systems is developed in \cite{GBYo2022}.
The approach developed here directly applies to non-isothermal versions of the diffusion through composite membranes presented in \cite{KeKa1963I}.

\section{Interconnected discrete-distributed systems}\label{section2}

We consider here the case of non-isothermal diffusion in a system made from discrete (well-stirred) as well as spatially distributed compartments. Such a situation is important in applications where homogenization techniques can be applied to some compartments to reduce the complexity of the system and the computational cost. For instance, in some studies of intracellular dynamics, the cellular and nuclear membranes must be considered as spatially distributed subdomains while homogenization techniques can be applied to the cytoplasm and the nucleus, see, e.g. \cite{Ch2012} and references therein.

\begin{figure}[h]
\begin{center}
\includegraphics[scale=0.4]{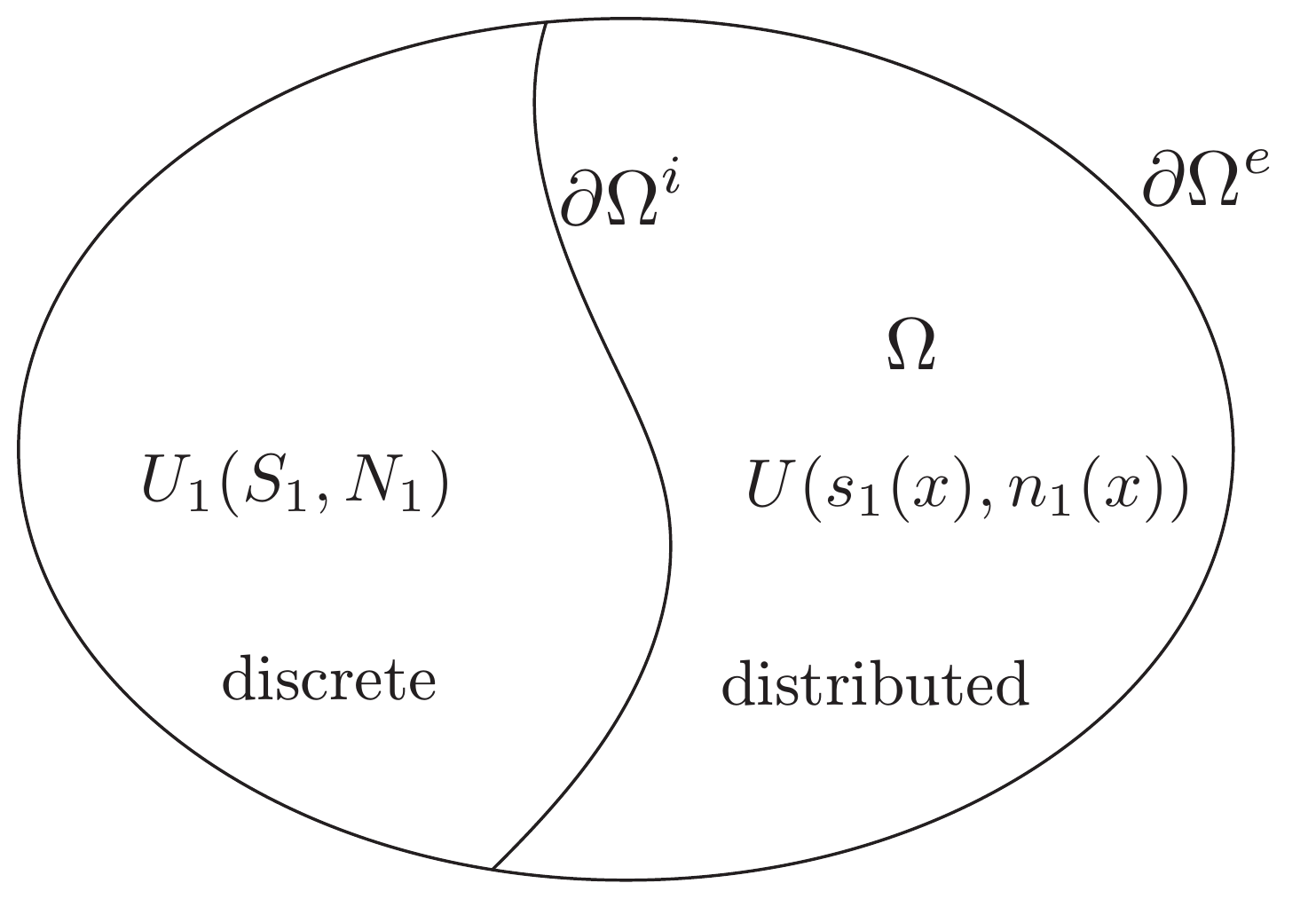}
\caption{\small The interconnected distributed-discrete system}
\label{discrete_distributed}
\end{center}
\end{figure}

For simplicity, we consider one well-stirred discrete compartment and one $n$-dimensional spatially distributed compartment, where the discrete compartment is assumed to have no interaction with the exterior. We denote by $ \Omega \subset \mathbb{R} ^n $, $n=1,2,3$ the domain of the well-stirred compartment, assumed to be compact with piecewise smooth boundary. The boundary $ \partial \Omega $ splits into two parts, namely the exterior boundary $\partial \Omega ^e$ and the interior boundary $\partial \Omega ^i$ which is in contact with the well-stirred compartment, see Fig. \ref{discrete_distributed}.

\subsection{Variational formulation}

Let us denote by $S$ and $N$ the entropy and the number of moles in the discrete compartment and by $s(x)$ and $n(x)$ the entropy density and mole number density in the distributed compartment $ \Omega $ with $ x \in \Omega $. We consider a general, possibly mixed discrete-distributed, Lagrangian function of the form $\mathscr{L}(S, N, s, n)$. We will denote by $\delta \mathscr{L}/ \delta s$ and $\delta \mathscr{L}/ \delta n$ the functional derivatives of $\mathscr{L}$ with respect to the density variables, defined by 
\begin{align*} 
&\left. \frac{d}{d\varepsilon}\right|_{\varepsilon=0} \mathscr{L}(S, N, s+ \varepsilon \delta s, n) = \int_ \Omega \frac{\delta \mathscr{L}}{\delta s} \delta s {\rm d} x,\\[2mm]
&\left. \frac{d}{d\varepsilon}\right|_{\varepsilon=0} \mathscr{L}(S, N, s, n+ \varepsilon \delta n) = \int_ \Omega \frac{\delta \mathscr{L}}{\delta n} \delta n {\rm d} x ,
\end{align*} 
for arbitrary $ \delta s_1$ and $ \delta n_1$. Such functional derivatives are assumed to exist.
If the Lagrangian is of the form
\begin{equation}\label{special_L} 
\mathscr{L}(S, N, s, n)= L(S, N) + \int_ \Omega \mathcal{L}(s, n) {\rm d} x
\end{equation} 
with $L(S,N)$ the Lagrangian of the discrete compartment and $\mathcal{L} (s,n)$ the Lagrangian density of the distributed compartment, then one has $ \delta \mathscr{L}/ \delta s= \partial \mathcal{L} /\partial  s$ and $ \delta \mathscr{L}/ \delta n= \partial \mathcal{L} /\partial  n$.

We denote by $ \mathbf{j}_s$ and $ \mathbf{j}$ and entropy flux and diffusive flux densities. By merging the variational setting for discrete thermodynamic systems (see \S\ref{section1} and \cite{GBYo2019}) and for continuum thermodynamic systems (see \cite{GBYo2016b,GBYo2019}), we get the following variational formulation
\begin{equation}\label{HP_mixed}
\begin{aligned} 
&\delta \int_{t_0}^{t_1} \!\Big[ \mathscr{L}(S, N, s, n) + \dot W N +\dot \Gamma  (S- \Sigma) \\
& \hspace{2cm} + \int_ \Omega \big(\partial _t w \,n + \partial _t  \gamma \,(s- \sigma )\big){\rm d} x \Big] {\rm d}t=0
\end{aligned}
\end{equation}
subject to the phenomenological constraints
\begin{equation}\label{phen_constraint_mixed}
\begin{aligned} 
\frac{\partial \mathscr{L}}{\partial S} \dot  \Sigma   &= \int_{ \partial \Omega ^i} \left( \dot \Gamma + \frac{\delta \mathscr{L}}{\delta s}  \biggr|_{ \partial \Omega ^i} \right) \, \mathbf{j} _s \cdot  \mathbf{n}\, {\rm d} \mathsf{A}  \\
& \qquad \qquad + \int_{ \partial \Omega ^i}\left( \dot W + \frac{\delta \mathscr{L}}{\delta n} \biggr|_{ \partial \Omega ^i}\right) \,\mathbf{j} \cdot \mathbf{n}\, {\rm d} \mathsf{A},   \\
\frac{\delta \mathscr{L}}{\delta s} \partial _t \sigma &=   \mathbf{j} _s \cdot \nabla \partial _t  \gamma  + \mathbf{j}\cdot \nabla \partial _t w,
\end{aligned}
\end{equation}
and the variational constraints
\begin{equation}\label{var_constraint_mixed}
\begin{aligned} 
\frac{\partial \mathscr{L}}{\partial S} \delta   \Sigma  &= \int_{ \partial \Omega ^i}   \delta  \Gamma \, \mathbf{j} _s \cdot  \mathbf{n}\, {\rm d} \mathsf{A} + \int_{ \partial \Omega ^i} \delta  W \, \mathbf{j} \cdot \mathbf{n}\, {\rm d} \mathsf{A},  \\[2mm]
\frac{\delta \mathscr{L}}{\delta s} \delta  \sigma &=  \mathbf{j} _s \cdot \nabla  \delta  \gamma    + \mathbf{j}\cdot \nabla \delta  w,
\end{aligned} 
\end{equation}
where $ \mathbf{n}$ denotes the outward pointing unit normal vector field to $ \partial \Omega$, $ {\rm d} \mathsf{A}$ is the area element on $\partial \Omega $, and where $ \delta W, \delta \Gamma , \delta w,  \delta \gamma$ vanish at $t=t_0,t_1$ and $ \delta w|_{ \partial \Omega }= \delta \gamma |_{ \partial \Omega }=0$. 
In the above, we introduce the thermodynamic displacements $W$, $\Gamma$ and   the thermodynamic displacement densities $w$, $\gamma$ and we also employ the internal entropy $\Sigma$ and the internal entropy density $\sigma$. One passes from the phenomenological constraints to the variational constraints by replacing the time rate of change of the thermodynamic displacements with $ \delta $-variations (such as $\dot \Gamma \leadsto \delta \Gamma $) and by removing the affine terms, as usual for constraints of thermodynamic type, \cite{GBYo2018a}.

\subsection{Equations of evolution}

Taking the variation of the integral in \eqref{HP_mixed} and using $ \delta W|_{t=t_0,t_1}=\delta \Gamma |_{t=t_0,t_1}=0$ and $ \delta w|_{t=t_0,t_1}= \delta \gamma |_{t=t_0,t_1}=0$, we get
\[
\begin{split}
&\int_{t_0}^{t_1} \Big[ \frac{\partial \mathscr{L} }{\partial N} \delta N +  \frac{\partial \mathscr{L} }{\partial S} \delta S + \delta N \dot W - \dot N \delta W  - ( \dot S- \dot \Sigma) \delta \Gamma + \delta S \dot \Gamma - \delta \Sigma \dot   \Gamma \\[2mm]
& \quad   + \int_ \Omega \Big(\frac{\delta \mathscr{L}  }{\delta s} \delta s +  \frac{\delta \mathscr{L}  }{\delta n} \delta n + \delta n \partial _t w - \partial _t n \delta w\\[2mm]
& \qquad\qquad\qquad - ( \partial _t s- \partial _t \sigma ) \delta \gamma + \delta s\partial _t \gamma - \delta \sigma \partial _t \gamma\Big){\rm d} x \Big] {\rm d} t=0.
\end{split}
\]
Since the variations $ \delta N$, $ \delta S$, $ \delta n$, and $ \delta s$ are free, one has
\[
\begin{split}
\delta N: & \;\;\frac{\partial \mathscr{L} }{\partial N} =- \dot   W,\quad   \;\;\;\,\delta S:\;\; \frac{\partial \mathscr{L} }{\partial S} = - \dot \Gamma,  \\[2mm]
\delta n: & \;\;\frac{\delta \mathscr{L}  }{\delta n} =- \partial _t  w,\quad  \;\;\delta s:\;\; \frac{\delta \mathscr{L} }{\delta s} = - \partial _t \gamma .
\end{split}
\]
From this, the previous condition becomes
\[
\begin{split}
&\int_{t_0}^{t_1} \Big[  - \dot N \delta W - ( \dot S- \dot \Sigma) \delta \Gamma + \delta \Sigma \frac{\partial \mathscr{L} }{\partial S} \\[2mm]
& +\int_ \Omega  \Big( - \partial _t n \delta w - ( \partial _t s- \partial _t \sigma ) \delta \gamma  +  \delta \sigma\frac{\delta \mathscr{L} }{\delta s} \Big) {\rm d} x \Big] {\rm d} t=0.
\end{split}
\]
Using now the variational constraint \eqref{var_constraint_mixed} and $ \delta w|_{ \partial \Omega }= \delta \gamma |_{ \partial \Omega }=0$, we get
\begin{align*} 
\delta W: & \;\; \dot  N = \int_{ \partial \Omega ^i } \mathbf{j} \cdot \mathbf{n} \,{\rm d} {\mathsf{A}}&  \;\;\delta  \Gamma:&\;\; \dot \Sigma =\dot S - \int_{ \partial \Omega ^i } \mathbf{j} _s \cdot \mathbf{n}\, {\rm d} \mathsf{A}\\[2mm]
\delta w: & \;\; \partial _t n + \operatorname{div} \mathbf{j} =0  & \;\;\delta \gamma :&\;\; \partial _t \sigma = \partial _t s + \operatorname{div} \mathbf{j} _s .
\end{align*}
To summarize we have obtained the evolution equations
\begin{equation}\label{equations}
\left\{
\begin{array}{l}
\vspace{0.3cm}\displaystyle\dot  N = \int_{ \partial \Omega ^i } \mathbf{j} \cdot \mathbf{n} \,{\rm d} {\mathsf{A}}, \qquad \partial _t n + \operatorname{div} \mathbf{j} =0\\
\vspace{0.3cm}\displaystyle\frac{\partial \mathscr{L}  }{\partial S} \Big( \dot S -   \int_{ \partial \Omega ^i }\!\! \mathbf{j} _s \cdot \mathbf{n}\, {\rm d} \mathsf{A} \Big) \\
\vspace{0.3cm}\displaystyle \quad = \int_{ \partial \Omega ^i}\!\! \Big(\! - \frac{\partial \mathscr{L}  }{\partial  S} +\frac{\delta \mathscr{L}   }{\delta s}\biggr|_{ \partial \Omega ^i} \Big) \, \mathbf{j} _s \cdot \mathbf{n}\, {\rm d} \mathsf{A} \\
\vspace{0.3cm}\displaystyle\qquad  + \int_{ \partial \Omega ^i}\Big( - \frac{\partial \mathscr{L}  }{\partial  N} +\frac{\delta \mathscr{L}   }{\delta n} \biggr|_{ \partial \Omega ^i}\Big) \,\mathbf{j} \cdot  \mathbf{n}\, {\rm d} \mathsf{A}\\[2mm]
\displaystyle\frac{\delta\mathscr{L}  }{\delta s} ( \partial _t s + \operatorname{div} \mathbf{j} _s) = - \mathbf{j} _s \cdot \nabla  \frac{\delta \mathscr{L}   }{\delta s}  - \mathbf{j} \cdot \nabla \frac{\delta \mathscr{L} }{\delta n}.
\end{array}
\right.  
\end{equation} 
%
When $\mathscr{L}$ is given by \eqref{special_L} with $L(S,N) = - U(S,N)$ the internal energy of the discrete compartment and $ \mathcal{L} (s, n)= - u(s(x),n(x))$ the internal energy density of the distributed compartment, we introduce $ T:=\dot\Gamma=-\frac{\partial \mathscr{L}}{\partial S}=  \frac{\partial U}{\partial S} $, $ \mathcal{M}:=\dot{W}=-\frac{\partial \mathscr{L}}{ \partial N} = \frac{\partial U}{\partial N}$,  $ \tau:=\partial_{t}\gamma=-\frac{\delta \mathscr{L}}{\delta s} = \frac{\partial u}{\partial s}$, $ \mu:=\partial_{t}w=-\frac{\delta \mathscr{L}}{\delta n} = \frac{\partial u}{\partial n}$. Hence, the last two equations in  \eqref{equations} read 
\begin{equation*}
\left\{
\begin{array}{l}
\vspace{0.2cm}\displaystyle T\Big( \dot S -   \int_{ \partial \Omega ^i } \mathbf{j} _s \cdot \mathbf{n}\, {\rm d} \mathsf{A}\Big)  = \int_{ \partial \Omega ^i} \Big( \tau|_{ \partial \Omega ^i}- T  \Big) \, \mathbf{j} _s \cdot \mathbf{n}\, {\rm d} \mathsf{A} \\
\vspace{0.3cm}\displaystyle \hspace{4cm}+ \int_{ \partial \Omega ^i}\Big( \mu |_{ \partial \Omega ^i}  - \mathcal{M}  \Big) \,\mathbf{j}\cdot \mathbf{n}\, {\rm d} \mathsf{A},\\
\displaystyle \tau ( \partial _t s + \operatorname{div} \mathbf{j} _s) = - \mathbf{j} _s \cdot \nabla  \tau - \mathbf{j} \cdot \nabla \mu .
\end{array}
\right.  
\end{equation*}
\subsection{Balance equations}

The mole balance of the distributed compartment can be written as
\[
\frac{d}{dt} \int_ \Omega n\, {\rm d} x= -\int_{ \partial \Omega ^i} \mathbf{j} \cdot \mathbf{n} \,{\rm d} \mathsf{A} - \int_{ \partial \Omega ^e} \mathbf{j} \cdot \mathbf{n} \,{\rm d} \mathsf{A}
\]
which shows the contributions associated with the exchanges across the interior and exterior boundaries. From this and the first equation in \eqref{equations}, the total mole balance reads
\[
\frac{d}{dt} \left[ N + \int_ \Omega n \,{\rm d} x \right] =  -\int_{ \partial \Omega ^e} \mathbf{j} \cdot \mathbf{n} \,{\rm d} \mathsf{A}.
\]

The entropy balances for the discrete and distributed compartments are found as
\[
\begin{split}
\dot S &=  \frac{1}{T}\int_{ \partial \Omega ^i} \Big( \tau|_{ \partial \Omega ^i}- T  \Big) \, \mathbf{j} _s \cdot \mathbf{n}\, {\rm d} \mathsf{A} \\
&\quad \underbrace{ \quad +  \frac{1}{T}\int_{ \partial \Omega ^i}\Big( \mu |_{ \partial \Omega ^i}  - \mathcal{M} \Big) \,\mathbf{j}\cdot \mathbf{n}\, {\rm d} \mathsf{A}}_{= \mathfrak{I}_{ \partial \Omega ^i }} + \underbrace{\int_{ \partial \Omega ^i } \mathbf{j} _s \cdot \mathbf{n}\, {\rm d} \mathsf{A}}_{= \mathcal{J} _S^{\rm int}}
\end{split}
\]
and
\[
\begin{split}
\frac{d}{dt} \int_ \Omega s \, {\rm d} x &= \underbrace{-\int_ \Omega \frac{1}{\tau} \,\mathbf{j} _s \cdot \nabla  \tau \,{\rm d} x -\int_ \Omega \frac{1}{\tau}  \,\mathbf{j} \cdot \nabla \mu \,{\rm d} x}_{= \mathfrak{I}_{ \Omega}} \\
&\quad -\underbrace{ \int_ { \partial \Omega^i }\mathbf{j} _s \cdot \mathbf{n} \,{\rm d} \mathsf{A}}_{=  \mathcal{J} _S^{\rm int}}-\underbrace{\int_ { \partial \Omega^e }\mathbf{j} _s \cdot \mathbf{n} \,{\rm d} \mathsf{A}}_{= - \mathcal{J} _S^{\rm ext}},
\end{split}
\]
where we denote by $\mathfrak{I}_{ \partial \Omega ^i }$ and $\mathfrak{I}_{ \Omega }$ the entropy generation rate for each compartment. We also denote by $ \mathcal{J} _S^{\rm int}$ the entropy flow rate  from the distributed to the discrete compartment and by $ \mathcal{J} _S^{\rm ext}$ the entropy flow rate  from the exterior to the distributed compartment. In addition, note that
\[
\mathfrak{I}_{ \partial \Omega ^i } = \dot \Sigma   \qquad\text{and}\qquad 
\mathfrak{I}_{ \Omega } = \frac{d}{dt} \int_ \Omega \sigma\, {\rm d} x 
\]
which show that the entropy generation rates coincide with the time rate of change of the variables $ \Sigma$ and $ \sigma $. This attributes a clear physical meaning to these two entropy variables appearing in the variational formulation \eqref{HP_mixed}-\eqref{var_constraint_mixed}.
The total entropy balance thus reads
\[
\frac{d}{dt} \left[  S +  \int_ \Omega s \, {\rm d} x \right] = \mathfrak{I}_{ \partial \Omega ^i }  + \mathfrak{I}_{ \Omega} +\mathcal{J} _S^{\rm ext} ,
\]
where the internal entropy flow rate $ \mathcal{J} _S^{\rm int}$ cancels out.

The energy balances for each compartment are 
\[
\begin{split}
\dot U &= \int_{ \partial \Omega ^i} \big( \tau|_{ \partial \Omega ^i } \mathbf{j} _s  + \mu |_{ \partial \Omega ^i }\mathbf{j} \big) \cdot \mathbf{n} \, {\rm d} \mathsf{A},\\
\frac{d}{dt} \int_ \Omega u \, {\rm d} x &= - \int_{ \partial \Omega } \big( \tau|_{ \partial \Omega  } \mathbf{j} _s  + \mu |_{ \partial \Omega }\mathbf{j} \big) \cdot \mathbf{n} \, {\rm d} \mathsf{A}
\end{split}
\]
so that the total energy balance is found as
\[
\frac{d}{dt} \left[ U+ \int_ \Omega u \, {\rm d} x   \right] =- \int_{ \partial \Omega ^e} \big( \tau|_{ \partial \Omega ^e } \mathbf{j} _s  + \mu |_{ \partial \Omega ^e }\mathbf{j} \big) \cdot \mathbf{n} \, {\rm d} \mathsf{A}= P^{\rm ext},
\]
where we recall that the boundary $ \partial \Omega $ of the distributed compartment splits into the internal $ \partial \Omega ^i $ and external $ \partial \Omega ^e$ boundaries and $P^{\rm ext}$ is the heat and matter power exchange from the exterior to the system. Since the discrete compartment is assumed to have no interaction with the exterior, the system is adiabatically closed if $ \mathbf{j}_s \cdot \mathbf{n} = \mathbf{j} \cdot \mathbf{n} =0$ on $ \partial \Omega ^e$.
The extension to the case where the discrete compartment can also exchanges heat and matter with the exterior can be achieved by appropriately combining the variational formulation \eqref{HP_mixed}--\eqref{var_constraint_mixed} with the approach developed in \cite{GBYo2018a}.

\subsection{Phenomenology}

From the second law and the expressions found for $\mathfrak{I}_ \Omega $ and $\mathfrak{I}_{ \Omega _i}$, the resulting form of entropy production suggests, in the linear regime, the phenomenological relations
\begin{equation}\label{pheno_open_system_1}
-\begin{bmatrix}
\vspace{0.2cm} \mathbf{j} _s\\
\mathbf{j} \\
\end{bmatrix}
=
\begin{bmatrix}
\vspace{0.2cm} \mathcal{L} _{ss} & \mathcal{L} _{sn}\\
 \mathcal{L} _{ns} & \mathcal{L} _{nn}\\
\end{bmatrix}
\begin{bmatrix}
\vspace{0.2cm} \nabla \tau\\
\nabla \mu \\
\end{bmatrix} \qquad  \text{on} \qquad   \Omega,
\end{equation}
and
\begin{equation}\label{pheno_open_system_2}
\begin{bmatrix}
\vspace{0.2cm} \mathbf{j} _s\! \cdot \mathbf{n} \\
\mathbf{j} \cdot \mathbf{n}  \\
\end{bmatrix}
=
\begin{bmatrix}
\vspace{0.2cm} \ell_{ss} & \ell _{sn}\\
\ell_{ns} & \ell_{nn}\\
\end{bmatrix}
\begin{bmatrix}
\vspace{0.2cm}  \tau|_{ \partial \Omega ^i}- T\\
 \mu |_{ \partial \Omega ^i}- \mathcal{M}\\
\end{bmatrix}\quad  \text{on} \quad  \partial \Omega ^i,
\end{equation}
for state functions $ \mathcal{L} _{ss},  \mathcal{L} _{sn},  \mathcal{L} _{ns},  \mathcal{L} _{nn}$ and $\ell_{ss},\ell_{sn}, \ell_{ns}, \ell_{nn}$ such that the symmetric parts of the $2 \times 2$ matrices are positive. 
In the diagonal case one gets the relations
\[
\begin{split}
&\mathbf{j} _s= - \mathcal{L} _{ss} \nabla T\text{ on }  \Omega, \, -  \mathcal{L} _{ss} \nabla T \cdot \mathbf{n} =\ell_{ss}\big(\tau|_{ \partial \Omega ^i}\!-\! T\big) \text{ on }  \partial \Omega ^i,\\[2mm]
&\mathbf{j} = - \mathcal{L} _{nn} \nabla \mu \text{ on } \Omega,   -  \mathcal{L} _{nn} \nabla n \cdot \mathbf{n} =\ell_{nn}\big( \mu |_{ \partial \Omega ^i} \!-\! \mathcal{M}\big) \text{ on }  \partial \Omega ^i,
\end{split}
\]
thereby giving Robin type of boundary conditions for heat and matter transfer through the internal boundary.

\subsection{One-dimensional case}

\begin{figure}[h]
\begin{center}
\includegraphics[scale=0.45]{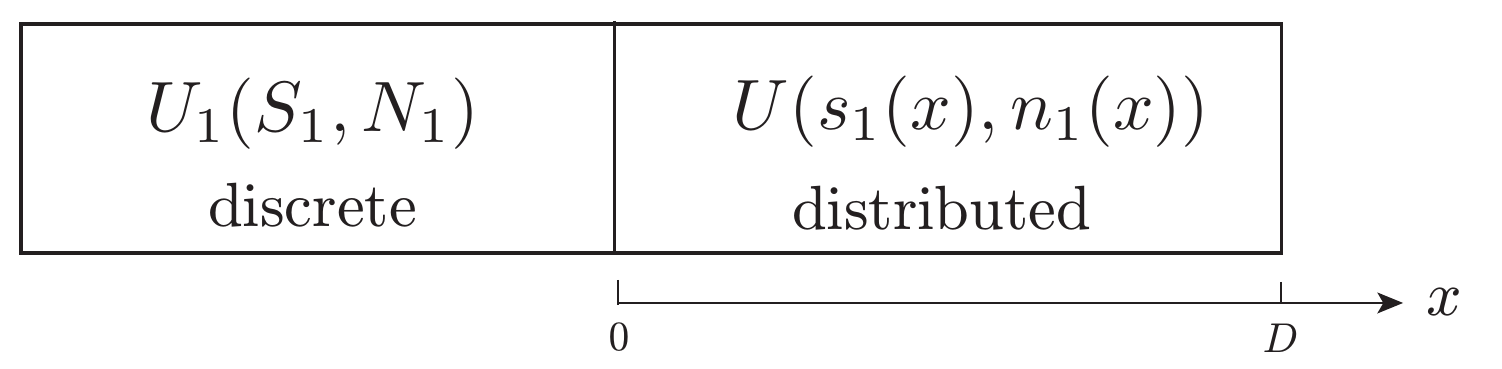}
\caption{\small The interconnected 1D distributed-discrete system}
\label{discrete_distributed_1D}
\end{center}
\end{figure}

As a simple instance of the variational approach developed in this section, we consider a one dimensional distributed compartment $ \Omega = [0,D] \subset \mathbb{R}$ as illustrated in Fig. \ref{discrete_distributed_1D}.

 The variational formulation reduces to
\begin{equation}\label{HP_mixed_1D}
\begin{aligned}
&\delta \int_{t_0}^{t_1} \!\Big[ \mathscr{L}(S,N, s, n)  + \dot W N +\dot \Gamma  (S- \Sigma )\\
& \qquad  \qquad + \int_0^D \big( \partial _t w \,n + \partial _t  \gamma \,(s- \sigma )\big){\rm d} x \Big] {\rm d}t=0
\end{aligned}
\end{equation}
subject to
\begin{equation}\label{phen_constraint_mixed_1D}
\begin{aligned} 
\frac{\partial \mathscr{L}}{\partial S} \dot  \Sigma   &=  -\left( \dot \Gamma +\frac{\delta \mathscr{L}}{\delta s} \biggr|_0 \right) \, j _s\biggr|_0 - \left( \dot W+\frac{\delta \mathscr{L}}{\delta n}  \biggr|_0\right) j\biggr|_0, \\[2mm]
\frac{\delta \mathscr{L}}{\delta s} \partial _t \sigma &= j_s   \partial _x   \partial _t \gamma +j \partial _x  \partial _t w ,
\end{aligned}
\end{equation}
and
\begin{equation}\label{var_constraint_mixed_1D}
\begin{aligned} 
\frac{\partial \mathscr{L}}{\partial S} \delta   \Sigma &  =    -\delta  \Gamma  j _s \bigr|_0 -\delta  W  j\bigr|_0, \\[2mm]
\frac{\delta \mathscr{L}}{\delta s} \delta  \sigma &= j_s \partial _x   \delta  \gamma + j \partial _x \delta  w,
\end{aligned} 
\end{equation}
with $ \delta w|_{ \partial \Omega }= \delta \gamma |_{ \partial \Omega }=0$, where $j_s|_0$ and $j|_0$ denotes the boundary values at $x=0$.
This formulation is obtained from \eqref{HP_mixed}--\eqref{var_constraint_mixed} by noting that the terms $\int_{ \partial \Omega ^i } \mathbf{j} \cdot \mathbf{n} \, {\rm d} \mathsf{A}$ and $\int_{ \partial \Omega ^e } \mathbf{j} \cdot \mathbf{n} \, {\rm d} \mathsf{A}$ correspond to $-j|_0$ and $j|_D$, respectively, similarly for $ \mathbf{j} _s$ and $j_s$.

For the Lagrangian \eqref{special_L} in which $L$ and $\mathcal{L}$ are given by the internal energies $U(S,N)$ and $u(s(x),n(x))$, we get from \eqref{HP_mixed_1D}--\eqref{var_constraint_mixed_1D} the evolution equations:
\begin{equation}\label{equations_1D}
\left\{
\begin{array}{l}
\vspace{0.2cm}\displaystyle\dot  N = - j|_0, \qquad \partial _t n + \partial _x j=0,\\[2mm]
\vspace{0.2cm}\displaystyle T\big( \dot S +  j_s|_0 \big)\! =\! -\big( \tau |_0- T   \big)j_s|_0- \big( \mu |_0 - \mathcal{M} \big)  j|_0,\\[2mm]
\displaystyle \tau ( \partial _t s + \partial _x j _s) \!=\! - j_s \partial _x   \tau - j \partial _x \mu .
\end{array}
\right.  
\end{equation} 
The various balances equations are easily deduced from this system.
In particular, the total energy and entropy balances take the following form
\begin{equation*}
\begin{split}
&\frac{d}{dt} \left[ S+ \int_ 0^D s \, {\rm d} x   \right]= - \frac{1}{T}\big(\tau|_0 - T\big) j_s|_0- \frac{1}{T}\big( \mu |_0 - \mathcal{M}\big) j_s|_0 \\
& \qquad   \quad  - \int_0^D \frac{1}{\tau} j _s\partial _x \tau \,{\rm d} x - \int_0^D \frac{1}{\tau} j \partial _x \mu\, {\rm d} x+ j_s|_D,\\[2mm]
&\frac{d}{dt} \left[ U+ \int_ 0^D u \, {\rm d} x   \right] =- \tau|_D j_s|_D -  \mu |_D j|_D.
\end{split}
\]
\vspace{-2mm}

\section{Interconnected distributed systems}\label{section3}
We consider here the case of a domain $ \Omega $ made from several interconnected distributed compartments $ \Omega _A$, $A=1,...,K$. We denote by $ \Sigma _{AB}= \partial \Omega _A \cap \partial \Omega _B$ the interface between compartments $A$ and $B$ and by $ \partial \Omega _A^e= \partial \Omega \cap \Omega _A$ the external boundary of compartment $ \Omega _A$.
Let $ \mathbf{j} _{A,s}$ and $ \mathbf{j} _A$ be the entropy and diffusive flux densities in compartment $A$ and let $ \mathbf{n} _A$ be the outward pointing unit normal vector field to $ \partial \Omega _A$.
The mass flux and energy flux continuity conditions across $ \Sigma  _{AB}$ imply
\begin{equation}\label{continuity_conditions}
\begin{aligned}
&(\mathbf{j} _A - \mathbf{j} _B) \cdot \mathbf{n} _A=0,\\[2mm]
&\big((T^A \mathbf{j} _{A,s} + \mu ^A \mathbf{j} _A) - (T^B \mathbf{j} _{B,s} + \mu ^B \mathbf{j} _B )\big) \cdot \mathbf{n} _A=0
\end{aligned} 
\end{equation} 
on $\Sigma _{AB}$.

For a given Lagrangian functional $\mathscr{L}(s_1,...,s_K, n_1,...,n_K)$, the continuum version of the variational formulation yields
\begin{equation}\label{HP_distributed}
\begin{aligned} 
&\delta \int_{t_0}^{t_1}  \Big[\mathscr{L}(s_1,...,s_K, n_1,...,n_K)  \\
& \; +\sum_A \int_ {\Omega_A}\big( \partial _t w^A n_A + \partial _t  \gamma ^A(s_A- \sigma _A)\big){\rm d}x \Big] {\rm d}t=0
\end{aligned}
\end{equation}
subject to
\begin{equation}\label{phen_constraint_distributed}
\begin{aligned} 
\frac{\delta  \mathscr{L}}{\delta s_A} \partial _t \sigma_A &= \mathbf{j} _{A,s} \cdot \nabla  \partial _t \gamma^A + \mathbf{j}_A \cdot \nabla \partial _t w ^A,
\end{aligned}
\end{equation}
and
\begin{equation}\label{var_constraint_distributed}
\begin{aligned} 
\frac{\delta \mathscr{L} }{\delta  s_A} \delta  \sigma ^A&= \mathbf{j} _{A,s} \cdot \nabla  \delta  \gamma ^A+ \mathbf{j}_A \cdot \nabla \delta  w^A, 
\end{aligned} 
\end{equation}
for $A=1,...,K$.

\subsection{Equations of evolution}

Since the computation of the critical condition is similar to the one presented above for the distributed compartment, we directly present the resulting conditions. We get
\[
\begin{split}
\delta n_A: & \;\;\frac{\delta \mathscr{L} }{\delta  n_A} =- \partial _t  w^A,  \;\;\delta s_A:\;\; \frac{\delta  \mathscr{L}}{\delta  s_A} = - \partial _t \gamma^A\\[2mm]
\delta w^A: & \;\; \partial _t n_A + \operatorname{div} \mathbf{j}_A =0,   \;\;\delta \gamma^A :\;\; \partial _t \sigma_A = \partial _t s_A + \operatorname{div} \mathbf{j} _{A,s},
\end{split}
\]
for $A=1,...,K$, so that the final system equations read
\begin{equation}\label{equations_distributed}
\left\{
\begin{array}{l}
\vspace{0.15cm}\displaystyle \partial _t n_A + \operatorname{div} \mathbf{j} _A=0 \\[2mm]
\vspace{0.15cm}\displaystyle \frac{\delta \mathscr{L} }{\delta s_A} ( \partial _t s_A \!+\! \operatorname{div} \mathbf{j} _{A,s}) \!=\! - \mathbf{j} _{A,s}\!\cdot\! \nabla  \frac{\delta \mathscr{L}}{\delta s_A} \! -\! \mathbf{j}_A \!\cdot \!\nabla \frac{\delta \mathscr{L} }{\delta n_A},
\end{array}
\right.  
\end{equation} 
for $A=1,...,K$.

For the Lagrangian functional $ \mathscr{L} (s_1,...,s_K,n_1,...,n_K)= -\sum_A\int_{ \Omega _A} u_A(s_A, n_A){\rm d} x$, with $u_A$ the internal energy density of compartment $A$, we have
\[
T^A\!:=\!-\frac{\delta \mathscr{L}}{\delta s_A}= \frac{\partial u_A}{\partial s_A} 
\;\;\textrm{and}\;\;
\mu ^A:=-\frac{\delta \mathscr{L}}{\delta n_A}=\frac{\partial u_A}{\partial n_A},
\]
so that the last equation becomes
\[
T^A( \partial _t s_A + \operatorname{div} \mathbf{j} _{A,s}) = - \mathbf{j} _{A,s}\cdot \nabla T^A - \mathbf{j}_A \cdot \nabla \mu ^A.
\]

\subsection{Balance equations}

The mole balance for compartment $ \Omega _A$ is
\[
\frac{d}{dt} \int_{ \Omega _A} n_A {\rm d} x = - \sum_B\int_{ \Sigma _{AB}} \mathbf{j}_A \cdot \mathbf{n} _A \, {\rm d} \mathsf{A} - \int_{ \partial \Omega ^e_A} \mathbf{j}_A \cdot \mathbf{n} _A \, {\rm d} \mathsf{A}, 
\]
which shows the contributions associated with the exchanges across $ \Sigma _{AB}$ and $ \partial \Omega _A^e$. 
From the first condition in \eqref{continuity_conditions}, the total mole balance reads
\[
\frac{d}{dt} \sum_A\int_{ \Omega _A} n_A {\rm d} x= - \sum_A\int_{ \partial \Omega ^e_A} \mathbf{j}_A \cdot \mathbf{n} _A \, {\rm d} \mathsf{A}.
\]

The total entropy balance is computed as
\begin{align*} 
&\frac{d}{dt}  \sum_A\int_ {\Omega_A} s_A {\rm d} x \\[2mm]
&= \sum_A\int_{ \Omega _A}\Big[ - \operatorname{div} \mathbf{j} _{s,A} - \frac{1}{T^A} \mathbf{j} _{s,A} \cdot \nabla T^A -\frac{1}{T^A} \mathbf{j} _{A} \cdot \nabla \mu ^A  \Big]{\rm d} x \\[2mm]
&= -  \sum_A\underbrace{\int_{ \Omega _A}\Big[ \frac{1}{T^A} \mathbf{j} _{s,A} \cdot \nabla T^A +\frac{1}{T^A} \mathbf{j} _{A} \cdot \nabla \mu ^A  \Big]{\rm d} x}_{= -\mathfrak{I}_{ \Omega _A}} \\
& \quad +\sum_{A<B} \underbrace{\int_{ \Sigma _{AB}} ( \mathbf{j} _{B,s}- \mathbf{j} _{A,s}) \cdot \mathbf{n} _A \, {\rm d} \mathsf{A}}_{\mathfrak{I}_{\Sigma _{AB}}}  -\sum_A \underbrace{\int_{\partial \Omega_A^e } \mathbf{j} _{A,s} \cdot \mathbf{n}_A\, {\rm d} \mathsf{A}}_{= -\mathcal{J} _S^{{\rm ext} \rightarrow A}},
\end{align*}
where $\mathfrak{I}_{ \partial \Omega_A }$ and $\mathfrak{I}_{ \Sigma _{AB} }$ are the entropy generation rate for each compartment and interface. We have also identified the entropy flow rate $ \mathcal{J} _S^{ {\rm ext}\rightarrow A}$ from  exterior to the distributed compartment $ \Omega _A $. We finally note the equality
\[
\mathfrak{I}_{ \Omega _A} = \frac{d}{dt} \int_ {\Omega _A} \sigma _A {\rm d} x,
\]
which relates $ \sigma _A$ to the entropy generation in $ \Omega _A$.
The energy balances for each compartment reads
\begin{align*} 
&\frac{d}{dt} \!\int_{ \Omega _A}\!\!u_A {\rm d} x = -\sum_B\int_{\Sigma _{AB}} \!\!\big( T^A|_{ \Sigma _{AB} } \mathbf{j} _{A,s} \! + \!\mu ^A |_{ \Sigma _{AB}  }\mathbf{j}_A \big) \!\cdot\! \mathbf{n}_A \, {\rm d} \mathsf{A}\\[2mm]
&  -\int_{ \partial \Omega _A^e} \big( T^A|_{ \partial \Omega _A^e} \mathbf{j} _{A,s}  + \mu ^A |_{\partial \Omega _A^e  }\mathbf{j}_A \big) \cdot \mathbf{n}_A \, {\rm d} \mathsf{A}\\[2mm]
&= \sum_B P^{B \rightarrow A} + P^{{\rm ext} \rightarrow A},
\end{align*}
where we have identified the power exchanges from $ \Omega _B$ to $ \Omega _A$ and from the exterior to $ \Omega _A$. Using the second condition \eqref{continuity_conditions} we have $P^{B \rightarrow A}=-P^{A \rightarrow B}$ so that the total energy balance is found as
\begin{align*} 
&\frac{d}{dt} \sum_A  \int_{\Omega_A} u_A \, {\rm d} x\\
& = -\sum_A \int_{ \partial \Omega _A^e} \big( T^A|_{ \partial \Omega _A^e} \mathbf{j} _{A,s}  + \mu ^A |_{\partial \Omega _A^e  }\mathbf{j}_A \big) \cdot \mathbf{n}_A \, {\rm d} \mathsf{A} = P^{\rm ext}.
\end{align*} 

\subsection{Phenomenology}
\vspace{-2mm}
From the second law and the expressions for $\mathfrak{I}_{ \partial \Omega_A }$ and $\mathfrak{I}_{ \Sigma _{AB} }$, we must have
\[
-\mathbf{j} _{s,A} \cdot \nabla T^A -  \mathbf{j} _{A} \cdot \nabla \mu ^A \geq 0 \quad\text{and}\quad ( \mathbf{j} _{B,s}- \mathbf{j} _{A,s}) \cdot \mathbf{n} _A\geq 0 .
\]
Using the two relations \eqref{continuity_conditions}, the second condition is 
\[
\big((T^A-T^B) \mathbf{j} _{A,s} + ( \mu ^A- \mu ^B) \mathbf{j} _A \big) \cdot \mathbf{n} _A\geq 0.
\]
This suggests, in the linear regime,  the phenomenological relations
\begin{equation}\label{pheno_distributed_1}
-\begin{bmatrix}
\vspace{0.2cm} \mathbf{j} _{A,s}\\
\mathbf{j}_A\\
\end{bmatrix}
=
\begin{bmatrix}
\vspace{0.2cm} \mathcal{L} ^A_{ss} & \mathcal{L} ^A_{sn}\\
 \mathcal{L} ^A_{ns} & \mathcal{L} ^A_{nn}\\
\end{bmatrix}
\begin{bmatrix}
\vspace{0.2cm} \nabla T^A\\
\nabla \mu^A \\
\end{bmatrix}
\end{equation} 
on $ \Omega  _A$, for all $A=1,...,K$ and
\begin{equation}
\begin{bmatrix}
\vspace{0.2cm} \mathbf{j} _{A,s}\! \cdot \mathbf{n}_A|_{ \Sigma _{AB}} \\
\mathbf{j}_A \cdot \mathbf{n}_A  |_{ \Sigma _{AB}}\\
\end{bmatrix}
=
\begin{bmatrix}
\vspace{0.2cm} \ell_{ss}^{AB} & \ell _{sn}^{AB}\\
\ell_{ns}^{AB} & \ell_{nn}^{AB}\\
\end{bmatrix}
\begin{bmatrix}
\vspace{0.2cm}  T^A- T^B\\
 \mu ^A- \mu ^B\\
\end{bmatrix}
\end{equation}
on $\Sigma _{AB}$, for all $A<B$, where the symmetric parts of the $2 \times 2$ are positive. As earlier, in the diagonal case one obtains Robin type interface conditions for heat and matter transfer.

\subsection{Future work}
\vspace{-2mm}
We project to analyze further how the variational formulation presented here  for the interconnected system can be systematically constructed from the variational formulation for each thermodynamic subsystem, in a similar way to the approach in \cite{JaYo2014} for interconnection in Lagrangian mechanics. 



\end{document}